\documentclass[acmsmall,screen,nonacm]{acmart}

\usepackage{booktabs}
\usepackage{graphicx}
\usepackage{float}
\usepackage{amsmath}
\usepackage{xcolor}
\usepackage{enumitem}
\usepackage{listings}
\usepackage{url}
\usepackage{tikz}
\usetikzlibrary{positioning, arrows.meta, calc, fit, shapes.geometric, backgrounds}

\setcopyright{none}
\ccsdesc[500]{Software and its engineering~Software notations and tools}
\ccsdesc[500]{Computing methodologies~Neural networks}
\ccsdesc[300]{Software and its engineering~Software testing and debugging}

\keywords{code language models, tensor product representations, interpretability, static analysis, causal interventions, program structure}

\newcommand{\Discover}{\textsc{Discover}}

\newcommand{\rqanswer}[2]{%
  \par\smallskip\noindent
  \fbox{\begin{minipage}{\dimexpr\linewidth-2\fboxsep-2\fboxrule\relax}
    \textbf{#1.}~#2
  \end{minipage}}\par\smallskip}

\begin{document}

\title{Program Bindings Inside Code Models}

\author{Prateek Kumar Rajput}
\affiliation{\institution{University of Luxembourg}\country{Luxembourg}}
\author{Yewei Song}
\affiliation{\institution{University of Luxembourg}\country{Luxembourg}}
\author{Jacques Klein}
\affiliation{\institution{University of Luxembourg}\country{Luxembourg}}
\author{Tegawend\'e F. Bissyand\'e}
\affiliation{\institution{University of Luxembourg}\country{Luxembourg}}

\begin{abstract}
Large language models are now used to complete, review, and repair
code, tasks that until recently relied on tools built on program
analysis. A refactoring engine, for example, resolves every use of a
name to its definition before it renames anything, and the
correctness of the rename depends on that information. A language
model exposes no such structure. Its state is a hidden vector per
token at each layer, and when it renames a variable in one place but
not in another, a developer cannot inspect which definition the
model associated with each use. In this paper, we test whether the
hidden states of a code model can be described and edited using the
information a compiler front end computes. For each token of a
Python program, standard analyzers provide its position in the
abstract syntax tree, its scope, its definition--use links, and its
syntactic category. We fit a decomposition that expresses every
hidden state of a frozen model as a sum of one term per token, each
combining the token with this structural label, and we replace the
model's own states with the decomposition at every layer while it
predicts the output of a small program. If the model still answers
correctly, the decomposition preserves the information the model
uses. We apply this test to $14$ open models, seven of them trained
for code, and compare against a baseline decomposition of the same
size built from token positions. The static-analysis decomposition
preserves the model's answers better than the positional one in $10$
of the $14$ models, including six of the seven code models, and
deleting the definition--use links from it removes most of that
advantage on the two models tested. Editing the decomposition, for
instance rebinding the argument of a call, makes StarCoder2-15B
produce the answer of the edited program in $90$ of $100$ cases,
compared with $19$ for the positional baseline. The Qwen2.5 family
is the exception on several measures. For most of the models tested,
standard static analysis therefore provides both a description of
what the hidden states encode about a program and a way to change it
without retraining.
\end{abstract}

\maketitle

\section{Introduction}
\label{sec:intro}

Large language models (LLMs) are now part of everyday software
development tools. They complete code in the editor, propose patches in code
review, and generate candidate fixes in automated program repair
\cite{fan2023llmse}. Welsh has argued that programming itself is on
its way to obsolescence, with software trained rather than written
\cite{welsh2023end}, and the observation that code is repetitive
enough to be predicted like natural language
\cite{hindle2012naturalness} explains much of why these models work.
Both are claims about what the models produce. What a model
internally represents about the program it is predicting remains an
open question. The tools that preceded these models rely on an
explicit representation of the program. A compiler resolves every
name to a definition before it generates code, a refactoring engine
renames a variable only after it has found every use that the
declaration binds, and a static analyzer reports a bug as a path
through definitions and uses. The correctness of these tools depends
on the representations they build. An LLM builds no representation
that a developer can inspect. It has no symbol table
and no scope resolver, only one hidden vector per token at each
layer, from which the next token is predicted. A recent survey
concludes that reliable systems will have to combine LLMs with
traditional program analyses \cite{fan2023llmse}, and such a
combination requires knowing how what the model represents relates
to what the analyses compute.

Figure~\ref{fig:example}(a) shows the kind of question we study. A
developer who adds a configuration to this command builder needs the
exact list the call returns in order to write the regression test,
which means following the selected variable through the call and
keeping the order of the arguments. Code models answer such
output-prediction questions well
\cite{nye2021scratchpads,liu2023codeexecution,gu2024cruxeval}, and
StarCoder2-15B \cite{lozhkov2024starcoder} is correct on $97.8\%$ of
the programs in our test set. When it is wrong, however, the
developer cannot inspect which list the model associated with
\texttt{arguments}, and when the model is asked to rename
\texttt{primary} throughout the program, nothing shows whether its
internal representation of the program changed accordingly. Studies
that perturb the input show the consequences. Neural models of code
change their predictions when variables are renamed
\cite{yefet2020adversarial,rabin2021generalizability}, and current
code models track data flow and control flow poorly under
counterfactual tests \cite{hooda2024counterfactual}. Whether a
model's hidden state is organized by the same bindings a compiler
tracks therefore determines how tools built on these models can be
checked. If it is, program analysis can be used to inspect, test,
and correct the model's internal state. If it is not, such tools can
only be validated through their outputs.

Most of the available evidence comes from probing. A classifier is
trained on frozen hidden states to predict syntactic or data-flow
properties, and its accuracy is taken as a measure of what the model
encodes
\cite{feng2020codebert,wan2022what,troshin2022probing,lpez2022astprobe,chen2022catprobing,shen2022benchmarking,velasco2024which,karmakar2023inspect,majdinasab2024deepcodeprobe,ahmed2023towards,nguyen2024empirical,ma2024unveiling}.
A probe shows that the information is present. It does not show that
the model uses it, because part of the accuracy can come from the
probe itself \cite{hewitt2019designing}, and it offers no way to
change what the model has encoded.

\begin{figure}[t]
\centering
\includegraphics[width=0.75\linewidth]{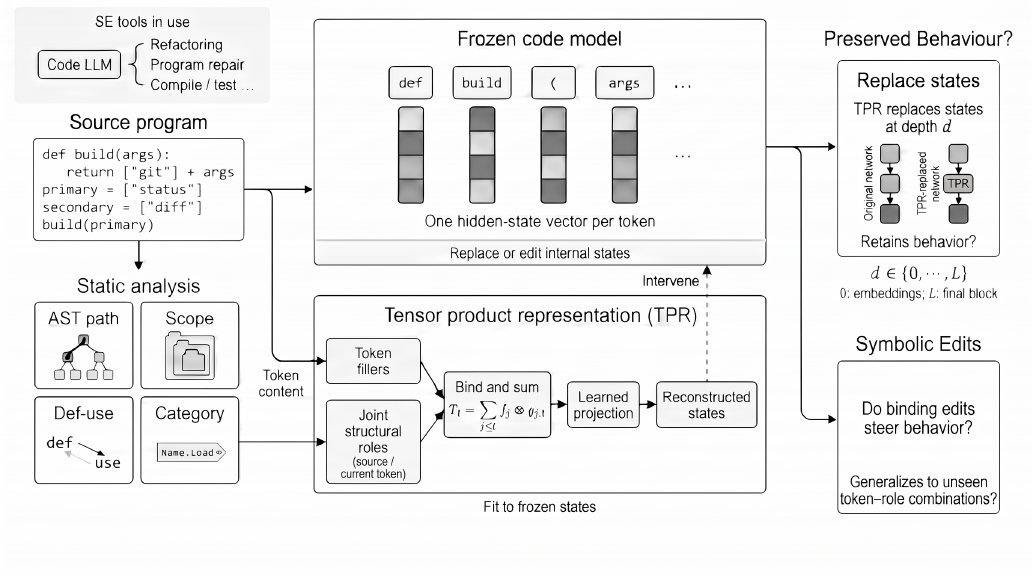}
\caption{Overview of the approach. Static analysis assigns each token
of the source program a structural label made of its AST path, scope,
definition--use record, and syntactic category. A tensor product
representation binds each token to its label, sums the terms, and
projects the sum to reconstruct the hidden states of a frozen code
model. The reconstruction is fitted to the model's own states and
then used in two behavioral tests. In the first, the model's states
are replaced by the reconstruction at all analyzed levels and we
check whether its answers are preserved, including on combinations of
tokens and roles never seen during fitting. In the second, the
bindings are edited, for instance by rebinding an argument or
renaming a variable, and we check whether the answer follows the
edit.}
\label{fig:overview}
\Description{Block diagram. On the left, a source program feeds a
static-analysis stage with four boxes labeled AST path, scope,
def-use, and category. In the center, a frozen code model shows one
hidden-state vector per token, and below it a tensor product
representation block takes token fillers and joint structural roles,
binds and sums them, applies a learned projection, and produces
reconstructed states that are fed back into the model as an
intervention. On the right, two boxes. The first, preserved
behavior, replaces states at a given depth and asks whether the
model retains its behavior. The second, symbolic edits, asks whether
binding edits change the behavior and whether the decomposition
generalizes to unseen token-role combinations.}
\end{figure}

In this paper, we take a different approach, summarized in
Figure~\ref{fig:overview}. We label every token of
a program with the information a standard Python analyzer computes
for it, namely its position in the abstract syntax tree, its
enclosing scope, its definition--use record, and its syntactic
category, obtained from the CPython parser, LibCST, and Beniget with
no parser written for the task. We then fit a decomposition that
expresses each hidden state of a frozen model as a sum of terms, one
per token, each combining the token with its structural label. Such
a sum is a tensor product representation in the sense of Smolensky
\cite{smolensky1990tpr}, in which the label is called the token's
role, and we fit it with the \Discover{} procedure of McCoy et al.\
\cite{mccoy2026emergent}. We call labels built from static analysis
SE roles and the baseline labels, which record only token positions,
positional roles. The test is behavioral. We substitute the
decomposition for the model's own states at every layer while the
model predicts a program's output, and we count how often the model
still produces the correct answer. Because the labels describe
program structure, the decomposition can also be edited in the same
way a refactoring tool edits source code. We can rebind the argument
of a call, or rename a variable, in the decomposition and check
whether the model's answer changes accordingly.

We apply this test to $14$ open models with 2B to 32B parameters.
Seven are trained for code, and six of them are paired with a
general-purpose model from the same family (Llama, Qwen2.5 at 3B and
at 32B, Gemma, DeepSeek, and Yi). The seventh pair, StarCoder2-15B
and Qwen2.5-14B, crosses families. The task is a set of small
list-manipulation programs whose expected outputs are obtained by
executing them, which allows us to withhold specific combinations of
values and positions from the fit and to align edits token by token.
The baseline is a decomposition of the same size built from
positional roles. The study addresses four research questions.
\begin{description}[leftmargin=3.2em,style=nextline,itemsep=1pt,topsep=2pt]
\item[RQ1] \emph{Does a decomposition built from SE roles preserve
the model's answers better than one built from positional roles?} If
it does, the information computed by program analysis describes the
model's internal state better than token positions do, which is a
prerequisite for using program analysis to inspect such models.
\item[RQ2] \emph{Does the fitted decomposition generalize to
combinations of values and positions it was never fitted on, and
which of its terms carry that generalization?} A decomposition that
only reproduces the combinations seen during fitting could not be
applied to new programs. The ablation of unseen terms shows whether
the roles themselves account for the generalization.
\item[RQ3] \emph{Do edits made on the decomposition change the
model's answer as the corresponding source edit would?} This
question is the most relevant one for tool support. If rebinding an
argument or renaming a variable in the model's state changes the
answer in the same way as the corresponding source edit, the state
can be edited and tested as a refactoring engine edits source code.
\item[RQ4] \emph{Is the effect stronger in models trained for code
than in their general-purpose counterparts?} A consistent difference
within the pairs would indicate that the structure is related to
training on code, with the caveat that the paired models differ in
more than their training data.
\end{description}

The results are consistent across five model families and reversed
in one. SE roles preserve the model's answers better than positional
roles in ten of the $14$ models, including six of the seven code
models, by $5.2$ to $38.3$ percentage points. On StarCoder2-15B the
decomposition preserves $79\%$ of the correct answers, compared with
$52\%$ for positional roles, and removing the definition--use
information from the labels costs $18.3$ and $28.9$ points on the two
models tested. The decomposition generalizes to withheld
combinations in two different ways, one for literal values and one
for variable names. Rebinding an argument in the decomposition makes
the model produce the new answer more often than the positional edit
does in all nine non-Qwen models, and when a rename breaks the
program so that its output changes, the model's answer changes
accordingly more often under SE roles in all $14$. Edits that change
a literal value do not benefit from SE roles. In six of seven pairs
the advantage is larger in the code model. The Qwen2.5
family is the exception on rebinding, on the variable-name
generalization, and on preservation below 32B. In all $14$ models
the SE decomposition reconstructs the hidden states less accurately
than the positional one, so selecting by reconstruction error alone
would have favored the wrong decomposition.

This paper makes three contributions.
\begin{itemize}[leftmargin=1.2em,itemsep=1pt,topsep=2pt]
\item A role scheme for decompositions of hidden states built from
standard static analyses, with no task-specific parser, and a
positional baseline of equal size against which it is measured.
\item A study of $14$ code and general-purpose models with the
controls needed to interpret the result, namely removal of the
definition--use field, an ablation that removes the terms absent
from the training data, control edits whose expected output changes,
and no-op and oracle-state checks.
\item A replication package with the code, the protocol and seeds,
the frozen corpora, and the per-program predictions behind every
table.
\end{itemize}
The study is limited to one synthetic task, one training seed per
condition, and a capped training budget, and the result depends on
the model family. The remainder of the paper is organized as follows.
Section~\ref{sec:background} presents the background and the running
example, Section~\ref{sec:approach} the role scheme,
Section~\ref{sec:method} the study design, Section~\ref{sec:results}
the results, and Sections~\ref{sec:discussion} to~\ref{sec:related}
the discussion, threats to validity, and related work.

\section{Background and Motivating Example}
\label{sec:background}

\subsection{Motivating Example}
\label{sec:example}

\begin{figure}[t]
\begin{minipage}[t]{0.5\linewidth}
\begin{lstlisting}[language=Python,numbers=none,xleftmargin=0pt,title={(a) A regression test for a command builder}]
def add_verbose(arguments):
    return arguments + ["--verbose"]
def build_command(arguments):
    return ["git"] + arguments
primary = ["status", "--short", "--branch"]
secondary = ["diff", "--stat", "--cached"]
build_command(primary)
# ['git', 'status', '--short', '--branch']
\end{lstlisting}
\end{minipage}%
\hfill
\begin{minipage}[t]{0.47\linewidth}
\begin{lstlisting}[language=Python,numbers=none,xleftmargin=0pt,title={(b) Test record \texttt{1f9387\ldots f6ab4f} of our held-out set}]
def alter(sequence):
    return sequence + ["A"]
def change(sequence):
    return ["E"] + sequence
y = ["C", "G", "K"]
x = ["P", "S", "X"]
change(y)
# ['E', 'C', 'G', 'K']
\end{lstlisting}
\end{minipage}
\caption{Two programs with the same binding structure. Program (a) is an
application example whose regression test asserts the exact list the
queried call returns. Program (b) is a measured test case of the
generalization experiment (RQ2). Under the renaming
\texttt{git}$\to$\texttt{E}, \texttt{status}$\to$\texttt{C},
\texttt{-{}-short}$\to$\texttt{G}, \texttt{-{}-branch}$\to$\texttt{K},
\texttt{diff}$\to$\texttt{P}, \texttt{-{}-stat}$\to$\texttt{S},
\texttt{-{}-cached}$\to$\texttt{X}, \texttt{-{}-verbose}$\to$\texttt{A},
\texttt{build\_command}$\to$\texttt{change},
\texttt{add\_verbose}$\to$\texttt{alter}, \texttt{primary}$\to$\texttt{y},
\texttt{secondary}$\to$\texttt{x}, and \texttt{arguments}$\to$\texttt{sequence},
the two abstract syntax trees are identical and both programs produce
the corresponding output when executed.}
\label{fig:example}
\Description{Two short Python listings side by side. Left, a command
builder with two functions, add_verbose and build_command, two
argument lists named primary and secondary, and a queried call
build_command(primary) whose executed answer is the list git, status,
dash dash short, dash dash branch. Right, a held-out test record with
two functions, alter and change, two three-letter lists y and x, and a
queried call change(y) whose executed answer is E, C, G, K. A caption
gives the one-to-one renaming under which the two programs have the
same abstract syntax tree.}
\end{figure}

Figure~\ref{fig:example}(a) is the command builder of the
introduction. To predict the output of the queried call, one has to
resolve a sequence of bindings. The argument \texttt{primary} is
bound to the parameter \texttt{arguments} of \texttt{build\_command},
the function prepends \texttt{"git"} to that list, and the other
function and the other list are not involved. The token text alone
does not provide this information. The parameter \texttt{arguments} occurs in both
functions with different scopes, each list name occurs once as a
definition and at most once as a use, and the same call with
\texttt{secondary} returns a different list. A compiler front end
records, for each token, the AST node it belongs to, the scope in
which it appears, the definitions a use may refer to, and its
syntactic category. These four pieces of information form the labels
we assign to tokens in Section~\ref{sec:approach}.

Figure~\ref{fig:example}(b) is a test record from our generalization
experiment (RQ2). Under the renaming given in the caption, its
abstract syntax tree is identical to that of (a), so the two programs
have the same binding structure. Record (b) is a held-out case
because the letters \texttt{C}, \texttt{G}, \texttt{K} and
\texttt{P}, \texttt{S}, \texttt{X} occupy list positions they never
occupied in the programs used for fitting, although every letter and
every position occurs separately in those programs. Generalization
here means combining values and structural positions that were each
seen during fitting but never together, which is also what the
developer's new configuration requires. Program (a) illustrates the
application scenario and was not used in the experiments. Record (b)
is an actual test case.

\subsection{Tensor Product Representations and \Discover{}}
\label{sec:tpr}

A tensor product representation (TPR) encodes a symbolic structure
as a set of role--filler bindings. Each content element (the filler)
is paired with the structural position it occupies (the role), and
the vector for the whole structure is the sum of the corresponding
outer products \cite{smolensky1990tpr,dolan1989tensor}. When the
role vectors are linearly independent, the filler bound to any role
can be recovered exactly \cite{smolensky2016basic}. TPRs are one
approach to
the binding problem, the question of how a distributed representation
keeps track of which value goes with which variable, position, or
syntactic slot \cite{greff2020binding,huang2026formalizing}. Whether
transformers solve this problem with an implicit binding scheme is an
open question
\cite{mccoy2018rnns,huang2018tensor,schlag2018learning,wu2025transformers,dhayalkar2025attention}.

\Discover{} (DISsecting COmpositionality in VEctor Representations)
tests whether a trained model implements a hypothesized set of
bindings by checking the model's behavior
\cite{mccoy2018rnns,soulos2019discovering,mccoy2026emergent}. The
researcher specifies a role scheme, a function that assigns each token a
structural role label. For each representation level $\ell$,
\Discover{} learns filler embeddings, role embeddings, and a linear
projection such that the frozen hidden state $h_t^{(\ell)}$ at
position $t$ is approximated by a sum of filler--role outer products
over the prefix. The approximation is then substituted for the
model's own state, and the model must still produce the correct
output. A successful replacement shows that the hypothesized
structure captures the information the model uses, and it allows
targeted interventions on fillers and roles. McCoy et al.\ report
this for GPT-OSS-20b on six symbolic tasks, including Python code
execution, where $31$ intervention types average $0.903$ accuracy
\cite{mccoy2026emergent}. Not every network can be approximated in
this way, and whether a given one can is an empirical question. We
keep the experimental settings of that paper
(Section~\ref{sec:training}) and change the target models, the
roles, and the replacement protocol, adding a held-out identifier
condition and a dependency-removal control under a $20{,}000$-step
training budget.

\subsection{Formal Instantiation}
\label{sec:formal}

Let $x_j$ be the tokenizer piece at query position $j$ and $t$ the
position whose state is reconstructed. Each piece receives a fixed
label $s_j$ from program analysis or from the prompt template
(Section~\ref{sec:approach}). For each level $\ell$, learned
embedding tables map pieces and role keys to vectors,
\begin{equation}
F_\ell\colon x_j \mapsto f_j^{(\ell)} \in \mathbb{R}^{100},
\qquad
R_\ell\colon r_{j,t} \mapsto q_{j,t}^{(\ell)} \in \mathbb{R}^{50},
\label{eq:embed}
\end{equation}
and the normalized hidden state is approximated by
\begin{equation}
T_t^{(\ell)} = \sum_{\substack{j \le t \\ j \in Q}}
F_\ell(x_j)\, R_\ell(r_{j,t})^{\top} \in \mathbb{R}^{100 \times 50},
\qquad
\widehat{z}_t^{(\ell)} = W_\ell \operatorname{vec}\!\big(T_t^{(\ell)}\big) + b_\ell,
\label{eq:tpr}
\end{equation}
followed by the inverse normalization
$\widehat{h}_t^{(\ell)} = \mu_\ell + \sigma_\ell \odot
\widehat{z}_t^{(\ell)}$. The state at position $t$ is thus
reconstructed from one binding per query token up to and including $t$, each
pairing the token's content with a role that depends on $(j,t)$.
Here $Q$ is the query span (the varying part of the prompt, including
a trailing output cue and excluding the fixed few-shot prefix),
$W_\ell \in \mathbb{R}^{d_\ell \times 5000}$ and $b_\ell \in
\mathbb{R}^{d_\ell}$ are the reconstruction weights and bias, and the
target is the per-level, per-channel normalized state $z_t^{(\ell)} =
(h_t^{(\ell)} - \mu_\ell) \oslash \sigma_\ell$. We estimate
$\mu_\ell$ and $\sigma_\ell$ once from the query-token states of
$1{,}024$ programs sampled from the $200{,}000$-program training
corpus and keep them fixed during fitting. Only $F_\ell$, $R_\ell$,
$W_\ell$, and $b_\ell$ are optimized. The model weights and the
symbolic role rules stay fixed. All learned parameters are
level-specific, and the discrete role definition is shared across
levels. The two schemes we compare differ in one component, the role
key. The positional baseline, \texttt{bidirectional\_all}, uses
$r_{j,t}^{\mathrm{pos}} = (j,\; t-j)$ with $j$ indexed relative to
the query span (the name refers to the two coordinates, with no
access to future tokens). Our main scheme
replaces this key with one derived from static analysis.

\section{Role Schemes from Static Analysis}
\label{sec:approach}

A role scheme is a hypothesis about which structural positions are
relevant when a model combines information about a program. For code, the
roles must capture the structure that bindings depend on, that is,
where each name is defined, which occurrences refer to it, and which
syntactic position each token occupies. They must also be computable
by standard tools on any program, so that the hypothesis depends only
on standard program analysis rather than on the task or on a learned
probe. Our label meets both requirements with four fields from
standard static analysis.

\subsection{Four Structural Fields}
\label{sec:fields}

For every tokenizer piece $x_j$ at source position $j$, the analyzer
assigns a composite label
\begin{equation}
s_j = (A_j,\, S_j,\, D_j,\, C_j),
\label{eq:composite}
\end{equation}
with the following fields.
\begin{enumerate}[leftmargin=*,itemsep=1pt,topsep=2pt]
\item $A_j$ \textbf{(AST path)}: the full CPython AST field and index
path of the smallest node containing the token, for example
\texttt{Module.body[2].value.elts[0]}. The path records structural
order, so SE roles also carry positional information.
\item $S_j$ \textbf{(scope)}: the kind and structural location of the
enclosing lexical scope, from the LibCST~1.8.6 \texttt{ScopeProvider}.
This field records where an occurrence appears, which need not be
where the referenced variable is defined.
\item $D_j$ \textbf{(dependency)}: for a definition, its structural
location. For a use, the sorted set of possible defining locations and
the expression-input dependencies computed by Beniget~0.5.0, with
explicit markers for builtins and unbound names.
\item $C_j$ \textbf{(syntactic category)}: the AST node class, the
load or store context where present, and the literal kind for
constants, for example \texttt{Name.Load}, \texttt{Name.Store}, or
\texttt{Constant.str}.
\end{enumerate}

The dependency field is purely symbolic. Every location is an AST path, a use
may list several candidate definitions, and the records are serialized
into $D_j$ and combined with the other fields into one categorical
label before fitting. The analysis is static and involves no
execution. The role vectors are learned, while the recorded locations
and relations remain fixed. These links correspond to the data
dependences of classical analysis, which connect definitions to the
uses they may reach \cite{ferrante1987program}. We use the records directly and
build no program dependence graph. The labels contain no control-flow
or control-dependence information, and the category field is purely
syntactic, since we perform no type inference. In
Figure~\ref{fig:example}(b), the \texttt{y} of the queried call is a
module-scope \texttt{Name.Load} with a possible defining location at
the first assignment, so the identifier receives different labels at
its declaration and at its use, and the two parameters named
\texttt{sequence} are separated by the scope field.

\subsection{From Program Text to Token Labels}
\label{sec:tokenlabels}

The analyzer processes the whole visible program. For each source
character it finds the smallest containing AST node, taking the
deeper node when two spans coincide, and assigns that node's label to the tokenizer piece
through the piece's first overlapping code character, after aligning
the UTF-8 offsets of the AST with the Unicode offsets of the
tokenizer. When an identifier spans several tokenizer pieces, the
pieces at its boundaries can therefore lose identifier-specific
detail. Tokens outside the source program,
including the trailing output cue, receive template labels. The
decomposition sums over the prefix of the query, but the analysis
reads the whole program.

\subsection{The Joint Source--Current Role}
\label{sec:jointrole}

Our main scheme, \texttt{se\_composite\_pair}, defines the role of a
source token relative to the position being reconstructed,
\begin{equation}
r_{j,t}^{\mathrm{SE}} = \operatorname{joint}(s_j, s_t),
\label{eq:jointkey}
\end{equation}
an ordered, serialized pair of composite labels used as a single
categorical key. The role of the \texttt{y} in the queried call of
Figure~\ref{fig:example}(b) is therefore one key when the
reconstructed state belongs to the closing parenthesis and another
when it belongs to the output cue. We do not embed the four fields
separately and then add or concatenate their vectors, and the source
and current components share no parameters across keys. On the main task the
vocabulary contains $1{,}009$ joint roles under the StarCoder2
tokenizer and $1{,}093$ under the CodeLlama tokenizer, excluding
padding.

\subsection{Positional Baseline}
\label{sec:posbaseline}

The positional baseline \texttt{bidirectional\_all} keeps
Equation~\eqref{eq:tpr} unchanged and replaces only the role key with
the positional pair $(j, t-j)$. Both schemes have the same table
shapes, the same projection size, and the same training budget, so
any difference in behavior between the two schemes is due to the role
definition alone.

\subsection{Relation to \Discover{}}
\label{sec:discover-relation}

The tensor product form, the linear reconstruction, the
source--current conditioning, replacement at all levels, and the
target-minus-source edit are taken from \Discover{}
\cite{mccoy2026emergent}, whose code experiment already includes value
and argument-variable edits among its six edit types. Its roles for
code come from a parser written for that task, which labels function
and variable identity, declaration order, positions inside definitions
and lists, and query positions. This paper adds the role
construction, four fields computed by standard analyzers on any Python
program with no task-specific parser, the held-out identifier
condition, the ablation of terms absent from training, the consistent
and invalid rename and the operand-order edits, the set of $14$ code
and general-purpose models, and the dependency-removal control. The
comparison we report is SE roles against positional roles of equal
capacity. We did not run \Discover{}'s hand-designed code roles as a
third condition, so we make no claim about them. Three
limitations apply. Only the dependency field has been ablated, on two
models. The joint key is categorical, so we report coverage for every
held-out condition (on the main task it was complete for all $14$
models). Finally, the roles come from the analyzer and are not
learned, so the experiment tests whether the model's behavior depends
on this particular structure. It does not test whether the model's
own organization coincides with it.

\section{Study Design}
\label{sec:method}

This section describes the task, the models, the fitting procedure,
and the replacement protocol shared by the four research questions,
and then the measurements used to answer each question.

\subsection{Task and Data}
\label{sec:task}

The main task follows the five-operation list task of \Discover{}'s
code experiment, regenerated as a fixed set of programs. Each
program has the same structure as the program in
Figure~\ref{fig:example}(b). It contains two functions,
\texttt{change} and \texttt{alter}, each applying one of five
operations (reverse, sort, repeat two to four times, prefix a letter,
suffix a letter), two lists \texttt{x} and \texttt{y} of one to three
uppercase letters, and one queried call. Function order, declaration
order, list values, and the queried function and list vary. Expected
outputs are obtained by executing the program and are checked
independently against the selected operation. Seeds $922001$ to
$922003$ generate $200{,}000$ training, $1{,}000$ validation, and
$1{,}000$ test programs, and generated programs that duplicate (by
SHA-256 hash) an earlier program, a program of the fixed development
set, or a program of another split are discarded. A prompt consists
of a fixed few-shot example, the program, and an output cue, and the
same JSONL files and hash manifests are used for all models. An AST
check of all $202{,}000$ programs confirms that they contain no
branches, loops, short-circuit expressions, or comprehensions. Every function has one
return statement, and builtins such as \texttt{sorted} are treated as
opaque. Predicting a program's output is the task that CRUXEval
poses at larger scale \cite{gu2024cruxeval}. The controlled grammar
is what allows us to execute every program, to withhold chosen
combinations, and to align edit triples token by token.

\subsection{Models}
\label{sec:models}

We study seven code-specialized models and seven general-purpose
counterparts (Table~\ref{tab:rq4}). The primary pair, StarCoder2-15B
\cite{lozhkov2024starcoder} and the Qwen2.5-14B base model
\cite{qwen2024qwen}, compares two model families of similar size. The
other six pairs come from the same family. CodeLlama-13B
\cite{rozire2023code} is paired with its documented base model
Llama-2-13B \cite{touvron2023llama2}, Qwen2.5-Coder-3B with
Qwen2.5-3B and Qwen2.5-Coder-32B with Qwen2.5-32B
\cite{hui2024qwencoder,qwen2024qwen}, CodeGemma-2B
\cite{codegemma2024} with Gemma-2B \cite{gemma2024},
DeepSeek-Coder-6.7B \cite{guo2024deepseekcoder} with DeepSeek-LLM-7B
\cite{bi2024deepseek} (both base variants), and Yi-Coder-9B with
Yi-1.5-9B \cite{young2024yi}. Training data and procedures differ
within every pair, so no pair isolates code pretraining or parameter
count. The frozen model runs in bfloat16 and the TPR and optimizer
states in float32. The number of analyzed levels (the embedding
output plus the output of every transformer block) ranges from $19$
to $65$ depending on the model.

\subsection{Fitting, Stopping, and Checkpoint Selection}
\label{sec:training}

Fillers have dimension $100$, roles $50$, and the projected tensor
$5{,}000$. The batch size is $12$, with at most $20{,}000$ optimizer
steps per condition, about $1.2$ passes over the training set. Normalization statistics come from $1{,}024$ training
programs, and fitting minimizes the normalized hidden-state MSE with
Adam at an initial learning rate of $10^{-4}$. Reconstruction error
on the $1{,}000$ validation programs is checked every $200$ steps,
and answer accuracy on $250$ validation programs every $10{,}000$
steps, the latter only as a diagnostic. For RQ1, after $20$ checks
without improvement the learning rate is halved, the best weights are
reloaded, and the optimizer state is reset, and fitting stops before
the eleventh reduction or at the step cap. For RQ2 the corresponding
numbers are $5$ checks and three reductions. RQ1 selects the
checkpoint by validation MSE and RQ2 by in-distribution validation
MSE plus the penalty below. No checkpoint is selected by task
accuracy or by any out-of-distribution result. All RQ1 checkpoints
lie between steps $19{,}400$ and $20{,}000$ and all RQ2 checkpoints
at step $20{,}000$. Every condition uses one configuration seed,
$20260922$, so every result is a single-seed measurement. RQ2 adds
the dimension-wise $L_{2,1}$ penalty on the embedding tables
described in the \Discover{} paper, with weight $10^{-4}$, which
drives whole embedding dimensions toward zero. The released
\Discover{} code applies an L1 penalty instead. We follow the written
protocol and record the difference.

\subsection{Replacement Protocol}
\label{sec:replacement}

For each token whose state is replaced, the replacement is applied
at all analyzed levels in the same forward pass. The states of the
query, including the output cue, are replaced, and the fixed few-shot
prefix is left unchanged. Attention caches are rebuilt by the decoder after
each replacement, and the answer is then generated as usual.
Identity-replacement and zero-ablation controls are run before any
fitting and scoring. All $14$ RQ1 audits report zero dropped
contributions, so no RQ1 number is affected by unknown keys, and the
original model's accuracy is recorded without any minimum capability
threshold.

\subsection{RQ1: Preservation, Layer Diagnostic, and Dependency Control}
\label{sec:rq1method}

We fit the positional and SE schemes separately against the same
frozen model, select by validation MSE, and evaluate replacement at
all levels on the test set, which is never used during fitting,
recording original accuracy,
accuracy under replacement, agreement with the original outputs,
retention on originally correct cases, reconstruction error, and key
coverage. A per-level diagnostic replaces one level at a time, on
$125$ in-distribution test programs, at all $41$ StarCoder2-15B
levels and records
accuracy and mean normalized reconstruction MSE per level. Audits
confirm that these programs contain no unseen filler--role pair among
$1{,}597{,}194$ contributions under either scheme. The
dependency-removal control refits the SE scheme with the dependency
field deleted from both the source and the current label before the
role vocabulary is rebuilt from the training set, leaving AST path,
scope, and category. It is a fresh fit with the same corpus, splits,
normalization sample, dimensions, batch size, and step cap, evaluated
on the same test programs with the same original predictions and
normalization arrays, on CodeLlama-13B and StarCoder2-15B only.
Removing the field merges joint roles, from $1{,}093$ to $914$ keys
on CodeLlama-13B and from $1{,}009$ to $838$ on StarCoder2-15B, so
the control has fewer role parameters than the full scheme although
the vector dimensions are equal.

\subsection{RQ2: Fresh Fits with Held-Out Combinations}
\label{sec:rq2method}

We distinguish two kinds of novelty. An \emph{unknown atomic key} is
a token piece or role key that has no embedding in the training
vocabulary. The encoder drops that contribution and counts it, so
coverage is reported together with accuracy. An \emph{unseen
pairing} is a filler--role combination whose two embeddings exist
but which never occurred together in training, so its outer product
is well defined. Unknown keys were the main obstacle in our early
attempts on repository code. RQ2 studies unseen pairings, of which
Figure~\ref{fig:example}(b) is an example. Because the RQ1 fits
were trained on the combinations we want to withhold, RQ2 fits
a new SE-only decomposition for each frozen code model, with no
imported weights, optimizer state, normalization statistics, or
validation cache. The grammar is restricted (seed $2026092309$) to
one prefix function, one suffix function, two three-letter lists over
six distinct letters, and the names \texttt{x}, \texttt{y},
\texttt{u}, \texttt{v}, with its own $200{,}000$ training, $1{,}000$
validation, and $1{,}000$ test programs and no code shared with
other splits or with the main corpus. List slots $0$ to $2$ lie in
the first declaration and $3$ to $5$ in the second. The test set
withholds the letters C/D, G/H, K/L, O/P, S/T, and W/X at slots $0$
to $5$, validation withholds A/B, E/F, I/J, M/N, Q/R, and U/V, and
every withheld letter occurs in other slots during training. The
identifier condition is separate. Training allows \texttt{y} or
\texttt{v} in the first declaration and \texttt{x} or \texttt{u} in
the second, validation uses \texttt{u} then \texttt{v}, and test uses
\texttt{x} then \texttt{y}, a new combination of names and
declaration slots that each occur in training. Identifier tests keep literal bindings allowed in
training, and literal tests keep identifier bindings allowed in
training. The evaluation covers $1{,}000$ in-distribution test
programs, $500$ literal held-out programs in which all six literal
bindings are withheld, and $1{,}000$ identifier held-out programs.
Held-out validation sets ($100$ and $250$ cases) are scored after
fitting, with novel pairs disjoint from the test set. Checks for
each tokenizer verify the expected outputs, disjoint hashes, known
atomic keys, and the absence of the withheld pairs from training.

\paragraph{Ablation of unseen terms.}
Let $\mathcal{S}$ be the set of filler--role pairs present in
training. We compare the full reconstruction with
\begin{equation}
T_{t,\mathrm{drop}}^{(\ell)} =
\sum_{\substack{j \le t,\ j \in Q \\ (x_j, r_{j,t}) \in \mathcal{S}}}
F_\ell(x_j)\, R_\ell(r_{j,t})^{\top},
\label{eq:dropnovel}
\end{equation}
which removes only the unseen contributions and keeps weights,
inputs, and known contributions unchanged. If accuracy falls, the
newly composed contributions are necessary for successful
replacement on the measured programs. The check does not reveal the
model's own algorithm, and an increase in accuracy is reported as
such. Under uniform guessing, the chance level is $(n-k)!/n!$, that
is $1/120 \approx 0.83\%$ for six letters with three returned, and
$1/2$ for the identifier swap. These chance levels are reference
points only, since a system with structural information can exceed
them. No positional and no general-purpose RQ2 fit was run.

\subsection{RQ3: Edits and Control Edits}
\label{sec:rq3method}

RQ3 uses the RQ1 checkpoints and triples of independently executed
programs (source, target, control). The text given to the model
remains the source program. For aligned token positions, the edit
adds the full target-minus-source difference of the decomposition to
the source states,
\begin{equation}
\Delta_t^{(\ell)} = \sigma_\ell \odot
\big(\widehat{z}_{t,\mathrm{target}}^{(\ell)} -
\widehat{z}_{t,\mathrm{source}}^{(\ell)}\big),
\qquad
h_{t,\mathrm{edited}}^{(\ell)} = h_{t,\mathrm{source}}^{(\ell)} + \Delta_t^{(\ell)}.
\label{eq:edit}
\end{equation}
The model reads the source text while its states carry the fitted
difference between the decomposition of the target and that of the
source. The per-level mean $\mu_\ell$ cancels in the difference.
These are whole-program differences, including changed roles, rather
than edits of a single SE field in isolation. A \emph{target hit} means that the edited model
produces the target program's executed output. The control condition
applies the control-minus-source difference instead and scores the
result against the control program's own executed output. A
\emph{control hit} therefore records that the control edit succeeded,
and it does not mean that the model was misled.

\begin{lstlisting}[language=Python,numbers=none,xleftmargin=0pt,caption={Source program of an edit triple. The queried call executes to \texttt{['Y', 'X']}.},label={lst:edit}]
def alter(sequence):
    return sequence + ["X"]
def change(sequence):
    return ["X"] + sequence
y = ["Y"]
x = ["Z", "U"]
alter(y)
\end{lstlisting}

Four edit families are measured with $100$ triples per model and
family. In terms of Listing~\ref{lst:edit}, an \emph{argument
rebinding} target calls \texttt{alter(x)}, with output
\texttt{Z,U,X}, and its control calls \texttt{change(y)}, with output
\texttt{X,Y}. A \emph{consistent rename} target swaps the \texttt{x}
and \texttt{y} declarations and calls \texttt{alter(x)}, which
preserves the output \texttt{Y,X}. Its control, an \emph{invalid
rename}, swaps the declarations but keeps \texttt{alter(y)}, which
changes the output to \texttt{Z,U,X}. An \emph{operand-order} target
makes \texttt{alter} prepend instead of append, with output
\texttt{X,Y}, and its control changes the unused function to a suffix
operation. A \emph{value change} alters a literal. Because a
consistent rename preserves the output, a target hit alone could mean
that the model simply kept its previous answer. The invalid-rename
control distinguishes following the edit from keeping the old
answer.
Triples require supported vocabulary, equal token counts, prefix
alignment, and distinct programs, and they are selected without
looking at model predictions. Unsupported cases are excluded rather
than counted as failures, and the operand-order family is excluded
for all five Qwen models because their candidates fail the
equal-alignment requirement. We report counts over all $100$
triples. Results restricted to the triples whose source and target
the original model answers correctly are in the replication package. No-op and oracle-state
controls check the intervention code.

\subsection{RQ4: Paired Model Contrasts}
\label{sec:rq4method}

For each model $m$ and test program $i$, let
$d_i^{(m)} = \mathbf{1}[\text{SE correct}] -
\mathbf{1}[\text{positional correct}]$. For each pair we compare
$\overline{d}^{(\mathrm{code})} - \overline{d}^{(\mathrm{general})}$
on identical test programs by paired resampling of the RQ1 results.
No new fitting is involved. RQ4 compares SE advantages rather than
raw accuracies, so a code model can show the larger advantage while
its counterpart has the higher accuracy, and because training data and
procedures differ within every pair, the result is a paired contrast between
specific models rather than an isolated effect of code pretraining.

\subsection{Statistics}
\label{sec:stats}

RQ1, RQ3, and RQ4 intervals use $5{,}000$ paired percentile bootstrap
resamples over the $1{,}000$ test programs with seed $20260923$. RQ2
uses the fixed-seed paired bootstrap of its saved evaluator. All
intervals are pointwise, exploratory, and uncorrected for
multiplicity. They measure test-case uncertainty under one fitting
seed and say nothing about variation across independent fits.

\subsection{Protocol Amendments}
\label{sec:amendments}

The study departed from the planned protocol in six respects, listed
once here. Section~\ref{sec:threats} states their consequences.
(1)~The $20{,}000$-step cap was adopted after interim validation and
before the final RQ1 test. Fits that stop at the cap have not
converged, and both schemes faced the same cap. (2)~The
StarCoder2-15B RQ1 SE fit was continued from an earlier schedule
change, whereas every other fit, including its dependency-removal
control, started fresh. A matched fresh full-SE reference was planned
and cancelled before it started, so the control comparison for
StarCoder2-15B differs in more than the removed field. (3)~The Yi-1.5-9B RQ1 fit was resumed
at step $10{,}000$, with optimizer and schedule state preserved,
after a device migration, which can change floating-point results.
(4)~A hardcoded $90\%$ original-validation threshold had excluded
Qwen2.5-Coder-3B ($223/250$), Qwen2.5-3B ($182/250$), and
CodeLlama-13B ($193/250$) from RQ2. We removed the threshold on
24~September~2026, since it selected models by capability, and kept
the weaker models in every table. (5)~The RQ2 penalty is $L_{2,1}$ as in
the \Discover{} paper's written protocol, whereas its reference code
applies L1. (6)~Direct GPU-to-GPU copies were found to corrupt
values, so all transfers are staged through the CPU and the affected
preliminary outputs were discarded. Fits run on rented GPUs (the 32B
Qwen pair, DeepSeek, and Yi-Coder) were imported with checksum
verification, so every reported number is backed by raw data held
locally.

\section{Results}
\label{sec:results}

All confidence intervals are paired bootstrap estimates
(Section~\ref{sec:stats}). Unless stated otherwise, accuracy is the
fraction of the $1{,}000$ test programs for which the model produces
the exact expected output.

\subsection{RQ1: Preservation under Replacement at All Levels}
\label{sec:rq1results}

\begin{table}[t]
\centering
\caption{RQ1. Accuracy after replacement at all levels on the
$1{,}000$ test programs, grouped by pair with the code-specialized
model first. ``Original'' is the unmodified model, ``Positional''
and ``SE'' the two schemes fitted against it. The gap is SE minus
positional in percentage points, with paired 95\% bootstrap
intervals for the two primary models. ``MSE'' is the validation
reconstruction error of the selected checkpoint, averaged over
levels.}
\label{tab:rq1}
\footnotesize
\setlength{\tabcolsep}{4pt}
\begin{tabular}{lrrrrrrr}
\toprule
Model & Levels & Original & Positional & SE & SE $-$ pos. & MSE pos. & MSE SE \\
\midrule
StarCoder2-15B & 41 & 97.8 & 52.0 & 79.0 & $+27.0$ $[23.8,30.2]$ & 0.08260 & 0.12129 \\
Qwen2.5-14B & 49 & 94.9 & 55.7 & 53.3 & $-2.4$ $[-5.8,+0.9]$ & 0.08037 & 0.10935 \\
\addlinespace
CodeLlama-13B & 41 & 80.5 & 36.4 & 61.9 & $+25.5$ & 0.06945 & 0.11234 \\
Llama-2-13B & 41 & 50.4 & 25.8 & 40.3 & $+14.5$ & 0.05321 & 0.08910 \\
\addlinespace
DeepSeek-Coder-6.7B & 33 & 89.6 & 44.0 & 64.9 & $+20.9$ & 0.07625 & 0.11638 \\
DeepSeek-LLM-7B & 31 & 44.2 & 23.2 & 19.9 & $-3.3$ & 0.04471 & 0.08340 \\
\addlinespace
Yi-Coder-9B & 49 & 95.7 & 43.5 & 81.8 & $+38.3$ & 0.10649 & 0.14561 \\
Yi-1.5-9B & 49 & 96.3 & 40.6 & 71.4 & $+30.8$ & 0.08442 & 0.12026 \\
\addlinespace
CodeGemma-2B & 19 & 33.1 & 12.9 & 18.1 & $+5.2$ & 0.05562 & 0.09043 \\
Gemma-2B & 19 & 44.1 & 22.1 & 24.1 & $+2.0$ & 0.04119 & 0.06672 \\
\addlinespace
Qwen2.5-Coder-3B & 37 & 87.1 & 49.3 & 28.8 & $-20.5$ & 0.06891 & 0.09538 \\
Qwen2.5-3B & 37 & 72.6 & 35.2 & 20.7 & $-14.5$ & 0.06277 & 0.08732 \\
\addlinespace
Qwen2.5-Coder-32B & 65 & 99.7 & 66.7 & 77.1 & $+10.4$ & 0.08613 & 0.12062 \\
Qwen2.5-32B & 65 & 99.6 & 66.1 & 73.2 & $+7.1$ & 0.08732 & 0.12041 \\
\bottomrule
\end{tabular}
\end{table}

Table~\ref{tab:rq1} gives the main result. The SE decomposition
preserves more correct answers than the positional decomposition in
$10$ of the $14$ models. Among the seven code-specialized models the
advantage holds in six, from $+5.2$ points on CodeGemma-2B to $+38.3$
on Yi-Coder-9B, and the exception is Qwen2.5-Coder-3B at $-20.5$.
Among the general-purpose models, Yi-1.5-9B, Llama-2-13B,
Qwen2.5-32B, and Gemma-2B favor SE roles, while Qwen2.5-14B ($-2.4$
$[-5.8,+0.9]$), Qwen2.5-3B ($-14.5$), and DeepSeek-LLM-7B ($-3.3$)
favor positional roles. The three Qwen2.5 models below 32B are
therefore the consistent exception, and the 32B Qwen pair favors SE
roles. The last two columns show that the SE fit has the higher
validation reconstruction error in all $14$ models, by $0.025$ to
$0.043$, including the ten in which it preserves more answers. Lower
reconstruction error therefore does not imply better preservation of
the model's answers.

\begin{figure}[H]
\centering
\includegraphics[width=0.7\linewidth]{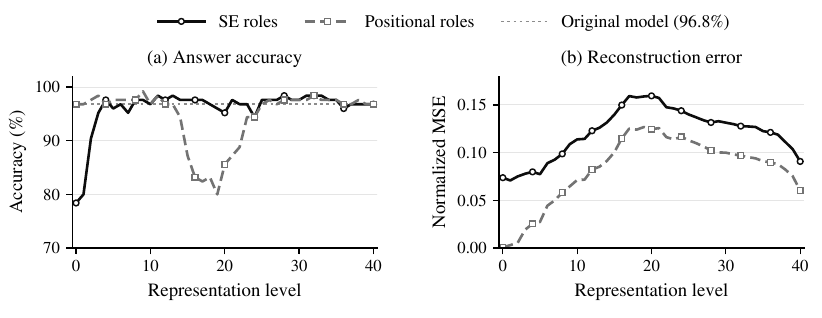}
\caption{Replacement of one level at a time on $125$ in-distribution
StarCoder2-15B test programs. (a)~Exact-answer accuracy, with the
unmodified model at $96.8\%$. (b)~Normalized reconstruction MSE,
weighted by query-token count. Level $0$ is the embedding output and
level $40$ the final normalized state.}
\label{fig:layer-diagnostic}
\Description{Two line plots over representation levels 0 to 40.
SE roles use a solid black line with circles, and positional
roles use a dashed gray line with squares. The accuracy panel includes
the original model's 96.8 percent reference line. Positional accuracy
falls to 80 percent at level 19, where SE accuracy is 96 percent.
Both schemes reach 96.8 percent at level 40. In the reconstruction
panel, SE has higher normalized MSE than positional roles at every level.}
\end{figure}

\paragraph{Per-level diagnostic.}
Figure~\ref{fig:layer-diagnostic} shows the same effect within a
single model, level by level. The unmodified StarCoder2-15B answers
$121$ of these $125$ programs correctly. Positional replacement preserves $121/125$ at the embedding
level against $98/125$ for SE, the order is reversed at level~19
($100$ against $120$), and both reach $121/125$ at level~40. The
positional scheme loses most at levels 15 to 21, where SE accuracy
stays between $95.2\%$ and $97.6\%$. SE nevertheless has the
larger mean normalized reconstruction error across the sweep
($0.12099$ against $0.08271$), and its MSE curve lies above the
positional one at every level. None of the $1{,}597{,}194$
contributions in these programs is unseen, so the diagnostic is
strictly in-distribution.

\paragraph{Dependency-removal control.}
The dependency field records candidate definition sites, which
describe possible value flow (Section~\ref{sec:fields}) and could be
regarded as semantic information inside the role definition. We
therefore remove the field and ask whether the remaining AST path,
scope, and category labels keep an advantage over positional roles.
Refitting the SE scheme without the dependency field lowers accuracy
from $61.9\%$ to $43.6\%$ on CodeLlama-13B ($-18.3$ points,
$[-20.9,-15.8]$) and from $79.0\%$ to $50.1\%$ on StarCoder2-15B
($-28.9$, $[-31.9,-26.1]$). On identical test programs, $197$
CodeLlama-13B programs and $293$ StarCoder2-15B programs are correct
only under the full scheme, against $14$ and $4$ correct only under
the control. The control remains above the positional scheme on
CodeLlama-13B ($43.6$ against $36.4$) and is at the positional level
on StarCoder2-15B ($50.1$ against $52.0$). Two qualifications
apply. The control has a smaller joint-role vocabulary ($914$ against
$1{,}093$ and $838$ against $1{,}009$ keys), and the StarCoder2-15B
full fit was continued from an earlier schedule while its control
started fresh (Section~\ref{sec:amendments}).

\rqanswer{RQ1}{SE roles preserve more answers than positional roles
in $10$ of $14$ models, including six of seven code-specialized
models, although they reconstruct states less accurately in all
$14$. Removing the dependency field costs $18.3$ and $28.9$ points.}

\subsection{RQ2: Held-Out Combinations and the Ablation of Unseen Terms}
\label{sec:rq2results}

\begin{table}[t]
\centering
\caption{RQ2. Fresh SE fits for the seven code-specialized models on
the restricted grammar. ``Orig.'' is the unmodified model, ``Full''
replaces all states with the complete decomposition, and ``Drop''
additionally removes the contributions whose filler--role pair was
absent from training (Equation~\ref{eq:dropnovel}). ``Full $-$
Drop'' carries a paired 95\% bootstrap interval. Chance levels are
$1/120 \approx 0.83\%$ for literals and $1/2$ for identifiers.}
\label{tab:rq2}
\footnotesize
\setlength{\tabcolsep}{2pt}
\begin{tabular}{lrrrlrrrl}
\toprule
& \multicolumn{4}{c}{Unseen literal bindings ($n{=}500$)} & \multicolumn{4}{c}{Unseen identifier bindings ($n{=}1{,}000$)} \\
\cmidrule(lr){2-5} \cmidrule(lr){6-9}
Model & Orig. & Full & Drop & Full $-$ Drop & Orig. & Full & Drop & Full $-$ Drop \\
\midrule
StarCoder2-15B      & 100.0 & 100.0 & 0.0 & $+100.0$ $[100.0,100.0]$ & 99.1  & 99.5 & 99.9  & $-0.4$ $[-0.8,-0.1]$ \\
CodeLlama-13B       & 96.4  & 83.2  & 0.0 & $+83.2$ $[80.0,86.4]$    & 95.8  & 94.8 & 95.4  & $-0.6$ $[-1.8,+0.5]$ \\
DeepSeek-Coder-6.7B & 99.0  & 90.2  & 0.0 & $+90.2$ $[87.4,92.6]$    & 97.9  & 94.0 & 91.2  & $+2.8$ $[+1.5,+4.1]$ \\
Yi-Coder-9B         & 99.8  & 100.0 & 0.0 & $+100.0$ $[100.0,100.0]$ & 100.0 & 99.5 & 100.0 & $-0.5$ $[-1.0,-0.1]$ \\
CodeGemma-2B        & 33.6  & 13.4  & 0.0 & $+13.4$ $[10.6,16.6]$    & 31.8  & 5.3  & 13.0  & $-7.7$ $[-9.6,-5.9]$ \\
Qwen2.5-Coder-3B    & 97.6  & 44.8  & 0.0 & $+44.8$ $[40.6,49.4]$    & 93.8  & 0.5  & 0.7   & $-0.2$ $[-0.5,+0.0]$ \\
Qwen2.5-Coder-32B   & 100.0 & 100.0 & 0.0 & $+100.0$ $[100.0,100.0]$ & 100.0 & 0.5  & 0.2   & $+0.3$ $[+0.0,+0.7]$ \\
\bottomrule
\end{tabular}
\end{table}

Table~\ref{tab:rq2} reports the held-out results of the fresh fits.
On the in-distribution test set of the restricted grammar, the fresh
fits preserve $88.1$ to $99.8\%$ of the answers on five models
(original accuracies $96.4$ to $100\%$), but only $27.4\%$ on
Qwen2.5-Coder-3B and $16.0\%$ on CodeGemma-2B (originals $97.0\%$
and $32.8\%$), which should be kept in mind when reading the
held-out results.

On held-out literal bindings the full decomposition solves $100\%$ of
the $500$ programs on StarCoder2-15B, Yi-Coder-9B, and
Qwen2.5-Coder-32B, $90.2\%$ on DeepSeek-Coder-6.7B, $83.2\%$ on
CodeLlama-13B, $44.8\%$ on Qwen2.5-Coder-3B, and $13.4\%$ on
CodeGemma-2B, against a chance level of $1/120$. Removing the newly
composed contributions drops every model to $0\%$. Where all $500$
paired differences are equal, the bootstrap interval collapses to
$+100$ points, which describes this sample only. In all seven models,
successful replacement on held-out literal bindings depends on the
newly composed filler--role contributions.

Held-out identifier bindings give a different result. On
StarCoder2-15B, Yi-Coder-9B, CodeLlama-13B, and DeepSeek-Coder-6.7B
the full decomposition solves $94.0$ to $99.5\%$ of the programs, and
removing the unseen contributions leaves that accuracy unchanged
($99.9$, $100.0$, $95.4$, and $91.2\%$, differences between $-0.6$
and $+2.8$ points). The known structural roles already carry the
binding information that the held-out identifiers would have
supplied. On both Qwen2.5-Coder models the full decomposition solves
only $0.5\%$ of the identifier programs, although the original models
solve $93.8\%$ and $100\%$ and although the 32B fit solves every
held-out literal program. On CodeGemma-2B it solves $5.3\%$, and
removing the unseen contributions \emph{raises} this to $13.0\%$
($-7.7$ $[-9.6,-5.9]$). The measurements do not establish why the
Qwen reconstructions fail on identifier recombination while
succeeding on literal recombination (Section~\ref{sec:disc-buys}).

\rqanswer{RQ2}{Held-out literal bindings are solved only through
newly composed contributions ($0\%$ without them in all seven
models). Held-out identifier bindings are solved with or without
them on four models and not at all on both Qwen2.5-Coder models.}

\subsection{RQ3: Edits on the Decomposition}
\label{sec:rq3results}

\begin{table}[t]
\centering
\caption{RQ3. Target hits, positional / SE, out of $100$ triples per
family, and invalid-rename control hits out of $100$. A target hit
means that the edited model produces the target program's output, a
control hit that the control edit produces the control program's
changed output. ``---'' marks the operand-order family, excluded for
the Qwen models by the equal-alignment requirement.}
\label{tab:rq3}
\footnotesize
\setlength{\tabcolsep}{5pt}
\begin{tabular}{lccccc}
\toprule
Model & Value & Rename & Argument & Operand & Invalid-rename control \\
\midrule
StarCoder2-15B      & 80 / 67 & 91 / 94 & 19 / 90 & 88 / 87 & 7 / 88 \\
Qwen2.5-14B         & 86 / 55 & 90 / 90 & 23 / 8  & ---     & 10 / 25 \\
\addlinespace
CodeLlama-13B       & 50 / 46 & 77 / 79 & 17 / 63 & 71 / 67 & 10 / 65 \\
Llama-2-13B         & 45 / 38 & 53 / 55 & 16 / 51 & 48 / 43 & 8 / 51 \\
\addlinespace
DeepSeek-Coder-6.7B & 62 / 64 & 88 / 88 & 10 / 73 & 68 / 77 & 4 / 78 \\
DeepSeek-LLM-7B     & 32 / 15 & 35 / 43 & 26 / 49 & 52 / 42 & 22 / 50 \\
\addlinespace
Yi-Coder-9B         & 71 / 82 & 95 / 95 & 2 / 88  & 88 / 90 & 3 / 92 \\
Yi-1.5-9B           & 84 / 70 & 91 / 96 & 7 / 84  & 74 / 82 & 4 / 84 \\
\addlinespace
CodeGemma-2B        & 28 / 5  & 17 / 22 & 7 / 23  & 28 / 24 & 8 / 29 \\
Gemma-2B            & 30 / 7  & 33 / 40 & 22 / 39 & 31 / 19 & 13 / 34 \\
\addlinespace
Qwen2.5-Coder-3B    & 84 / 13 & 88 / 86 & 37 / 13 & ---     & 18 / 33 \\
Qwen2.5-3B          & 54 / 11 & 70 / 71 & 35 / 21 & ---     & 15 / 29 \\
\addlinespace
Qwen2.5-Coder-32B   & 94 / 64 & 99 / 98 & 55 / 7  & ---     & 31 / 44 \\
Qwen2.5-32B         & 93 / 62 & 98 / 99 & 39 / 3  & ---     & 14 / 37 \\
\bottomrule
\end{tabular}
\end{table}

Table~\ref{tab:rq3} summarizes the edits for all $14$ models.
Argument rebinding gives the clearest result, and it separates the
models in the same way as RQ1. The SE edit produces the target answer
more often than the positional edit in all nine non-Qwen models, with
$19$ against $90$ hits on StarCoder2-15B, $2$ against $88$ on
Yi-Coder-9B, $10$ against $73$ on DeepSeek-Coder-6.7B, $17$ against
$63$ on CodeLlama-13B, and $7$ against $84$ on Yi-1.5-9B, and with
the smallest margins on the Gemma pair. In all five Qwen models the SE
edit does worse ($23$ against $8$ on Qwen2.5-14B, $55$ against $7$ on
Qwen2.5-Coder-32B). On the StarCoder2-15B triples whose source and
target the original model answers correctly, the counts are $19/95$
for the positional edit and $87/95$ for the SE edit, a difference of
$+71.6$ points $[62.1,81.1]$.

Under a consistent rename, both schemes keep the output in most
models, and the SE edit matches or exceeds the positional edit in $12$
of $14$ models (the two Qwen2.5-Coder models are lower by two and by one hit). Because a consistent
rename preserves the output, the invalid-rename control is the
informative measurement. Its output changes, and the SE edit produces
the changed output more often than the positional one in all $14$
models, with $88$ against $7$ on StarCoder2-15B, $92$ against $3$ on
Yi-Coder-9B, $78$ against $4$ on DeepSeek-Coder-6.7B, and $65$
against $10$ on CodeLlama-13B. The smallest margin is $44$ against
$31$ on Qwen2.5-Coder-32B. The rename results therefore reflect the
direction of the edit rather than a tendency to keep the old answer.
Value changes favor the positional edit in $12$ of $14$ models, often
by a wide margin ($84$ against $13$ on Qwen2.5-Coder-3B), with
DeepSeek-Coder-6.7B ($62$ against $64$) and Yi-Coder-9B ($71$ against
$82$) as the exceptions. Operand-order edits give mixed results on
the nine models with eligible triples (SE ahead on three). No-op and
oracle-state controls passed for all models.

\rqanswer{RQ3}{SE edits produce the target answer for argument
rebinding more often than positional edits in all nine non-Qwen
models and less often in all five Qwen models, and they follow the
invalid-rename control more often in all $14$. Value changes favor
the positional edit in $12$ of $14$.}

\subsection{RQ4: Paired Model Contrasts}
\label{sec:rq4results}

\begin{table}[t]
\centering
\caption{RQ4. Difference between the SE advantage (SE minus
positional accuracy, Table~\ref{tab:rq1}) of each code-specialized
model and that of its general-purpose counterpart, in percentage
points, with paired 95\% bootstrap intervals on identical test
programs.}
\label{tab:rq4}
\footnotesize
\setlength{\tabcolsep}{3.5pt}
\begin{tabular}{llrrl}
\toprule
Code / general pair & Relationship & Code $\Delta$ & General $\Delta$ & Difference [95\% CI] \\
\midrule
CodeLlama-13B / Llama-2-13B & derived from the base model & $+25.5$ & $+14.5$ & $+11.0$ $[+6.8,+15.2]$ \\
DeepSeek-Coder-6.7B / DeepSeek-LLM-7B & same developer, similar size & $+20.9$ & $-3.3$ & $+24.2$ $[+20.0,+28.7]$ \\
Yi-Coder-9B / Yi-1.5-9B & same developer and size & $+38.3$ & $+30.8$ & $+7.5$ $[+3.2,+11.7]$ \\
CodeGemma-2B / Gemma-2B & built on the base model & $+5.2$ & $+2.0$ & $+3.2$ $[+0.0,+6.4]$ \\
Qwen2.5-Coder-3B / Qwen2.5-3B & same family and size & $-20.5$ & $-14.5$ & $-6.0$ $[-9.8,-2.2]$ \\
Qwen2.5-Coder-32B / Qwen2.5-32B & same family and size & $+10.4$ & $+7.1$ & $+3.3$ $[-0.2,+7.0]$ \\
StarCoder2-15B / Qwen2.5-14B & different families & $+27.0$ & $-2.4$ & $+29.4$ $[+24.8,+34.0]$ \\
\bottomrule
\end{tabular}
\end{table}

Table~\ref{tab:rq4} pairs each code-specialized model with its
general-purpose counterpart. The code model has the larger SE
advantage in six of the seven pairs. The clearest comparison is
CodeLlama-13B against its documented base model Llama-2-13B, $+25.5$
against $+14.5$ points, a difference of $+11.0$ $[+6.8,+15.2]$. The
DeepSeek and Yi pairs give $+24.2$ $[+20.0,+28.7]$ and $+7.5$
$[+3.2,+11.7]$. The Gemma pair ($+3.2$ $[+0.0,+6.4]$) and the 32B
Qwen pair ($+3.3$ $[-0.2,+7.0]$) point in the same direction with
intervals that touch or include zero, and the cross-family primary
pair gives $+29.4$ $[+24.8,+34.0]$. The 3B Qwen pair is the exception
at $-6.0$ $[-9.8,-2.2]$, where both models favor positional roles and
the code model does so more. RQ4 compares advantages rather than raw
accuracies, so CodeGemma-2B has the larger advantage although its SE
accuracy ($18.1\%$) is below Gemma-2B's ($24.1\%$).

\rqanswer{RQ4}{The SE advantage is larger in the code-specialized
member of six of seven pairs, most clearly for CodeLlama-13B against
Llama-2-13B ($+11.0$ $[+6.8,+15.2]$). The 3B Qwen pair is the
exception ($-6.0$ $[-9.8,-2.2]$).}

\section{Discussion}
\label{sec:discussion}

\subsection{Dependence on the Model Family}
\label{sec:disc-modeldep}

The main finding depends on the model family. With the same protocol
for all models, SE roles produce the target answer for argument rebinding more often
than positional roles in all nine non-Qwen models and preserve
behavior better in eight of them (DeepSeek-LLM-7B, at $-3.3$, is the
exception), across five model families and sizes from 2B to 15B. In the
Qwen2.5 family the same roles perform worse on most measurements that
involve bindings. Three of the five Qwen models favor positional roles
for preservation, all five favor positional edits for argument
rebinding, and both Qwen2.5-Coder fits fail on held-out identifiers
while succeeding on held-out literals. The exception is partial,
since the 32B pair favors SE roles for preservation and the
invalid-rename control favors SE roles on every Qwen model, but every
binding-related measurement points the same way. The completed
experiments do not identify a cause. The Qwen models differ from the
others in tokenizer, training data, and training procedure, and the
measurements do not explain why their identifier reconstructions
fail while the original models solve the same programs. We therefore
interpret the Qwen results as a limitation of the \Discover{} method
with these roles rather than as evidence about what the Qwen models
encode, since an approximation can fail because of the role
definition, the optimization, or the decoder's sensitivity to
perturbation \cite{mccoy2026emergent,mccoy2026reconciling}.

Table~\ref{tab:rq1} also has a methodological consequence. The SE fit
reconstructs states less accurately than the positional fit in every
model and still preserves more answers in ten of them, so
reconstruction loss is a poor criterion for choosing between
structural hypotheses, and behavioral preservation under replacement
is the relevant criterion.

\subsection{Which Edits the Roles Support}
\label{sec:disc-buys}

The edit experiments distinguish two kinds of edit. Edits that change
the structure of the program succeed under SE roles, and edits that
substitute content do not. Argument rebinding moves a value from one
binding context to another, which is the operation that the scope
and dependency fields describe, and the SE edit is better on all nine
non-Qwen models. Consistent renaming is handled by both schemes, and
the invalid-rename control, which is followed more often under SE
roles on all $14$ models, shows that the SE-edited model responds to
the direction of the edit. Value changes favor the positional edit on
$12$ of $14$ models. A hypothesis consistent with this asymmetry is
that the joint source--current key captures the relation between a
source token and the reconstructed position, which rebinding and
renaming change, while value edits change content that the
decomposition represents only indirectly through the fillers.

The dependency-removal control indicates, on two models, where this
relational information comes from. Without the dependency field the
SE scheme loses $18.3$ points on CodeLlama-13B and $28.9$ on
StarCoder2-15B, where it falls to the level of the positional scheme.
The dependency field records which definition a use may refer to,
which is also what argument rebinding changes.

The held-out experiment adds a third observation. Generalization to
held-out literal bindings depends on newly composed filler--role
contributions in every model, whereas generalization to held-out
identifier bindings does not, on the four models that generalize at
all. For a program analysis, a constant bound to a list slot and an
identifier bound by a declaration are the same kind of binding, but
the fitted decomposition treats them differently. Measured by
accuracy alone, both cases would have counted as compositional
generalization, and
the Qwen identifier failure would have been hidden behind a literal
accuracy of $100\%$. One explanation consistent with the data is that
literal values are the contents that must appear in the answer, so
removing their new contributions removes information the answer
needs, whereas a variable name selects a binding without appearing in
the answer, so the remaining structural contributions can still
identify the selected binding. This explanation has not been tested.

\subsection{Scope of the Task and Implications}
\label{sec:disc-task}

The list-execution task is small by design. It provides an executed
expected output for every program, allows exact filler--role
combinations to be withheld while every atomic key stays in the
training vocabulary, and yields edit triples with exact token
alignment. None of these controls is available on repository code
today, where our early attempts failed on unknown atomic keys
(Section~\ref{sec:rq2method}). The task is nevertheless close to practice. Figure~\ref{fig:example}
shows a regression test with the same token-level structure as a
measured held-out record, and the held-out condition corresponds to
a developer rearranging a known configuration.

For tool builders, the result means that refactoring-like operations
can be applied to the model's state instead of its input, at least on
tasks where the fit is accurate. The invalid-rename control shows
that such edits act in the direction of the change, which is a
requirement for using them as counterfactual tests. Earlier work
showed, by modifying inputs, that models of code are sensitive to
renaming \cite{yefet2020adversarial,rabin2021generalizability} and
weak at tracking data and control flow \cite{hooda2024counterfactual}.
A fitted decomposition allows the same questions to be asked of the
model's internal state, one binding at a time. For research on what
code models encode, the result argues for complementing probing
\cite{wan2022what,karmakar2023inspect,majdinasab2024deepcodeprobe,ma2024unveiling}
with behavioral replacement tests and for treating the model family
as an experimental variable, since a result that held in five
families failed in a sixth. The difference between identifiers and
literals also cautions against general claims of compositional
generalization in code models
\cite{hajipour2024simscood,vegner2025behavioural}, since how a
decomposition generalizes depends on the kind of binding.

\section{Threats to Validity}
\label{sec:threats}

\paragraph{Construct validity.}
Our composite label is one way of representing software-engineering
structure. Only the dependency field has been ablated, on two models,
and the control has a smaller role vocabulary than the full scheme,
so the measured loss bounds the field's contribution without
isolating it. AST path, scope, and category are evaluated jointly.
The category field involves no type inference, and the dependency
field reports static may-definitions and expression inputs rather
than runtime data flow or control dependences, which the
straight-line grammar does not exercise. Token-level projection can lose detail on the boundary pieces of
multi-token identifiers, and the analysis reads the whole program. We
compare SE roles with positional roles only. \Discover{}'s
hand-designed code roles were not run, and the value and
argument-variable edit families originate in that work.

\paragraph{Internal validity.}
Every condition is a single-seed measurement (seed $20260922$), and
the sensitivity to the seed is unknown. The six amendments of
Section~\ref{sec:amendments} affect internal validity as follows.
The StarCoder2-15B SE fit was continued from an earlier schedule
while every other fit started fresh, so its RQ1 pairing and its
dependency control did not follow the same procedure. The Yi-1.5-9B fit
was resumed after a device migration. Fits that stop at the cap have
not converged, although both schemes faced the same cap. The RQ2
numbers correspond to the $L_{2,1}$ penalty and not to the L1
penalty of the reference code. Checkpoint selection used validation
MSE and never task accuracy. RQ3 excludes ineligible triples instead
of scoring them as failures, so the operand-order family covers nine
models, and the invalid-rename control is a single diagnostic per
model.

\paragraph{Conclusion validity.}
Bootstrap intervals are pointwise, exploratory, and uncorrected for
multiplicity, and they quantify test-case uncertainty under one
fitting seed only. Per-model RQ1 gaps carry intervals for the two primary models
only. Counts such as ``six of seven'' summarize directions across
heterogeneous models and are not tests of a common effect.

\paragraph{External validity.}
The task is a synthetic list-execution family with straight-line
programs, one return per function, opaque builtins, lists of at most
three single-letter elements, and a four-name identifier vocabulary,
in Python only. The results do not establish behavior on real
repositories. Figure~\ref{fig:example}(a) is an illustration, and an
earlier pilot on repository code was inconclusive. The $14$ models span six families and 2B
to 32B parameters, but every code/general pair differs in training
data and procedure, so RQ4 does not isolate code pretraining or
parameter count, and no general-purpose model has an RQ2 fit.

\section{Related Work}
\label{sec:related}

\subsection{What Code Models Encode}
\label{sec:probing-related}

Probing studies measured how well syntax, AST structure, and data
flow can be decoded from the hidden states of CodeBERT-family
encoders
\cite{feng2020codebert,wan2022what,troshin2022probing,lpez2022astprobe,chen2022catprobing,shen2022benchmarking,velasco2024which}
and, later, of decoder models
\cite{karmakar2023inspect,majdinasab2024deepcodeprobe,ahmed2023towards,nguyen2024empirical,ma2024unveiling},
coding agents \cite{silva2026latent}, and state-space code models
\cite{wu2026towards}. All of these measure decodability. Behavioral
studies instead change the input. Models of code change their
predictions under semantics-preserving transformations such as
renaming \cite{yefet2020adversarial,rabin2021generalizability},
counterfactual tests find that current code models track data flow
and control flow poorly \cite{hooda2024counterfactual}, and causal
interpretability estimates the effect of program properties on
predictions by causal inference over inputs
\cite{palacio2024causation}. Our study differs from all three lines,
since it replaces states at all levels of a frozen model with roles
from static analysis and intervenes on the states rather than on the
input.

\subsection{Tensor Product Representations and Binding}
\label{sec:tpr-related}

TPRs originate in cognitive modeling
\cite{smolensky1990tpr,dolan1989tensor} and have been used as
architectural components
\cite{huang2018tensor,schlag2018learning,smolensky2016basic} and as
hypotheses about trained networks, which recurrent networks implement
implicitly \cite{mccoy2018rnns} and role-learning networks can
discover \cite{soulos2019discovering}. The binding problem has a long
history in neural computation \cite{greff2020binding} and has
recently been formalized for modern architectures
\cite{huang2026formalizing}. Related analyses ask how transformers
learn variable binding in symbolic programs \cite{wu2025transformers},
read attention as a vector-symbolic operation
\cite{dhayalkar2025attention}, probe TPR-structured directions
\cite{lee2026tensor}, and connect filler--role structure with
mechanistic interpretability \cite{enyan2026unifying}. Whether
LLMs implement symbolic structure at all remains open
\cite{mccoy2026reconciling,mccoy2024when,gu2024cruxeval}, and
\Discover{}'s central result, frozen representations replaced by
closed-form symbolic expressions with behavior largely intact,
supports a quasi-symbolic view \cite{mccoy2026emergent}. RQ2 measures
representational systematicity \cite{vegner2025behavioural}
behaviorally, and the ablation of unseen terms is a counterpart of
the control tasks used for probes \cite{hewitt2019designing}. Our results extend the question to $14$
open models with roles from program analysis and add a boundary,
since the decomposition succeeds in five model families and fails
in the Qwen2.5 family under the same protocol.

\subsection{Causal Interventions on Internal States}
\label{sec:causal-related}

Our RQ3 edits are activation-level interventions. Causal abstraction aligns a network with a high-level
causal model through interchange interventions
\cite{geiger2021causal}, causal tracing patches activations between
clean and corrupted runs \cite{meng2022locating}, and
circuit-discovery methods localize computations in graphs of
components \cite{conmy2023towards,syed2023attribution,hadad2026formal}.
\Discover{} instead derives the edit from a fitted decomposition of
the target program and adds the full target-minus-source difference,
including changed current-token roles, and its code experiment
reports six edit types, including value and argument-variable edits
\cite{mccoy2026emergent}. We keep that design, add the rename and operand-order conditions
with their control edits, and run the edits on $14$ models.

\section{Conclusion}
\label{sec:conclusion}

We asked whether the hidden states of open code models can be written
as sums of bindings whose roles come from static analysis, whether
such decompositions generalize to held-out combinations, and whether
they support edits that act like refactorings on the model's state.
On $14$ models and a controlled list-execution task, the answers are
consistent in five model families and weaker or reversed in one. SE
roles preserve more answers than positional roles in $10$ models,
including six of the seven code-specialized models, and edits on the
decomposition produce the target answer for argument rebinding more
often than positional edits in all nine non-Qwen models. Within the
scope of one synthetic task, a single seed per condition, and a
capped training budget, roles from ordinary static analysis make a
code model's program state inspectable and editable without
retraining in most of the models tested, and not in one family.
Future work includes multi-seed replication, further ablations,
control flow in the task grammar, and repository programs.

\section*{Data Availability}
\label{sec:data-availability}

An anonymized replication package is publicly available at
\url{https://zenodo.org/records/23078805}.
It contains \texttt{se-tpr-code.zip}, released under the MIT license,
and \texttt{se-tpr-data-results.zip}, whose synthetic data and
experimental results are released under CC BY 4.0. The code includes
the role extraction, fitting and intervention implementations,
protocol configurations and pinned model revisions, corpus generators,
and scripts for verification and result analysis.

The data archive contains the frozen synthetic corpora and edit
triples, per-program predictions for the four research questions and
the two dependency-removal controls, and five CSV result tables.
SHA-256 manifests and provenance records support file verification.
The tables can be regenerated from the saved predictions on a CPU
without downloading or running the LLMs. Pretrained model
weights, fitted TPR checkpoints, optimizer states, and hidden-state
caches are excluded from the release. The supplied code supports fresh
GPU runs using the documented protocol and upstream model revisions.

Upon acceptance, we will release a non-anonymized version of the
package with author attribution.

\paragraph{AI assistance.}
Codex assisted with packaging the replication artifact and preparing
the plotting script for Figure~\ref{fig:layer-diagnostic} from saved
measurements. The plotted accuracies and token-weighted reconstruction
errors were checked against the per-program records.

\bibliographystyle{ACM-Reference-Format}
\bibliography{refs}

\end{document}